\documentclass[twocolumn,showpacs,aps,prl,superscriptaddress]{revtex4-1}
\usepackage{graphicx,amsmath,amssymb}
\usepackage{dcolumn,fancyhdr}
\usepackage{bm,natbib}
\usepackage{times}
\usepackage{txfonts}
\usepackage{epstopdf}
\usepackage{latexsym}
\usepackage{epsfig}

\usepackage{color}
\definecolor{Blue}{rgb}{0,0,1}
\definecolor{Red}{rgb}{1,0,0}
\definecolor{Green}{rgb}{0,1,0}
\definecolor{Purp}{rgb}{.2,0,.2}
\definecolor{white}{rgb}{1,1,1}

\begin{document}
\title{Unification of witnessing initial system-environment correlations and witnessing non-Markovianity}

\author{C\'esar A. Rodr\'iguez-Rosario}
\email{cesar.rodriguez@bccms.uni-bremen.de}
\affiliation{Bremen Center for Computational Materials Science, Universit\"at Bremen, Am Fallturm 1, 28359 Bremen, Germany}

\author{Kavan Modi}
\affiliation{Department of Physics, University of Oxford, Clarendon Laboratory, Oxford UK}
\affiliation{Centre for Quantum Technologies, National University of Singapore, Singapore}

\author{Laura Mazzola}
\affiliation{Centre for Theoretical Atomic, Molecular and Optical Physics, School of Mathematics and Physics, Queen's University Belfast, BT7 1NN Belfast, United Kingdom}

\author{Al\'an Aspuru-Guzik}
\affiliation{Department of Chemistry and Chemical Biology, Harvard University, Cambridge MA, USA}

\begin{abstract}
We show the connection between a witness that detects dynamical maps with initial system-environment correlations and a witness that detects non-Markovian open quantum systems. Our analysis is based on studying the role that state preparation plays in witnessing violations of contractivity of open quantum system dynamics. Contractivity is a property of some quantum processes where the trace distance of density matrices decrease with time. From this, we show how a witness of initial-correlations is an upper bound to a witness of non-Markovianity. We discuss how this relationship shows further connections between initial system-environment correlations and non-Markovianity at an instance of time in open quantum systems.
\end{abstract}

\maketitle

\section{Introduction}

The open quantum systems perspective, where a system interacts with an environment, is central to the study of dephasing, dissipation, and decoherence ~\cite{Sudarshan:1961p2145, *SudarshanJordan61, Angelbook}. Historically, the research has focused on open system dynamics that are Markovian, completely positive and have no initial system-environment correlations. There has been recent interest in studying quantum stochastic processes beyond these assumptions~\cite{Rodriguez08, Rodriguez11b, ApollaroPRA11, *Raj12a}. Many witnesses have been proposed to detect non-Markovianity~\cite{WolfPRL08, BreuerPRL09, RivasPRL10, Fisher, RajagopalPRA10, Haikka:2012tk}. Also, there has been proposals for witnessed that detect initial system-environment correlations~\cite{LaineEPL10, Gessner:2011un}, and have motivated experimental investigations~\cite{PhysRevA.84.032112}. The witness for these properties share many similar properties, but their connection has not been established. 

In this paper, we will show the essential relationship between witness of non-Markovianity and witnesses of initial system-environment correlations. We start with a brief review of the properties of open quantum systems. we then review some formulations of witnesses of non-Markovianity. We then review essential features of open systems with  initial system-environment correlations. We discuss the central role that state \emph{preparation}~\cite{preppap} plays when going beyond the assumptions of Markovianity and no initial system-environment correlations. Using this, we derive a witness of initial-correlation that is also an upper bound to a non-Markovianity witness. We then discuss the role of Markovian effects in the relationship between witnesses of initial correlations and witnesses of non-Markovianity.

We can represent quantum stochastic processes using dynamical maps. Dynamical maps are super-operators derived from the dynamics of the $SE$ state $\rho^{SE}$ that evolves through time due to some total unitary $U\equiv U_{t,\tau}= \mathcal{T}\exp{\left\{-i \int_t^\tau H^{SE}(t^\prime) \,dt^\prime\right\}}$. The dynamical map of a system state $S$ is 
\begin{equation}\label{map}
\rho^S(\tau)=\mbox{Tr}_E\left[U_{t,\tau} \rho^{SE}(t) U^\dagger_{t,\tau} \right]=\mathbb{B}_{t,\tau}\left(\rho^{S}(t) \right).
\end{equation}
Since the map transforms states from time $t$ to states at time $\tau$, the time $t$ is called the \emph{initial time}, where the process described by $\mathbb{B}$ starts. If there are no correlations between the system and the environment at the initial time $t$, such that $\rho^{SE}(t)= \rho^S(t) \otimes \rho^E$, the map is completely positive \cite{choi72a,choi75,Stinespring}. Completely positive maps are interesting because they have a simple mathematical structure \cite{choi72a,choi75}, and form a time-dependent semigroup, such that they have the composition of two completely positive maps $\mathbb{B}_{{t_1},{t_2}}\circ \mathbb{B}_{{t_0},{t_1}}$ is also a completely positive map. The simplicity of this structure makes the no initial-correlations assumption  appealing. However, recently there has been much interest in relaxing this assumption \cite{pechukas94a, *Jordan:2006p112, *jordan:052110, Rodriguez:2007p123, Rodriguez08, Rodriguez11b, embeddingmap, Hayashi, kuah:042113, buzek, Carteret08a, lidarZD, masillo:012101, Hutter2011, Modi2012a}. 

\section{Witnesses}

Recall that a witness $\mathcal{W}$ of some property $w$ is a mathematical object such that: If $\mathcal{W}> 0$, then the property $w$ is satisfied. Some of the non-Markovianity witnesses have focused on detecting deviations from completely positive dynamics as a feature of non-Markovianity called divisibility \cite{WolfPRL08, Hou:2011dd}. Others have focused on detecting deviations from contractive dynamics as witnesses of non-Markovianity \cite{WolfPRL08, BreuerPRL09, Breuer:2009cua, UshaDevi:2010wy, Fisher, RajagopalPRA10, Dijkstra:2011us, Kossakowski:2012wu}. Since completely positive maps are contractive, these witnesses share a lot of common features ~\cite{HaikkaPRA11, ChruscinskiPRA11}.

Note that Markovian quantum stochastic processes also form a time-dependent semigroup. No initial-correlations lead to completely positive maps, and completely positive maps form a time-dependent semigroup. This is why the concepts of complete positivity, no initial-correlation and Markovianiaty  are often tied all together in the literature \cite{Royer:1996vv, WolfPRL08, Andersson:2010wr, ChruscinskiPRA11, Kossakowski:2012wu, Dijkstra:2011us}.  Although the formulation of witnesses of non-Markovianity and of initial-correlation have similar features, their relationship has not been studied.  In this paper, we derive an unified approach that relates a class of witnesses of initial-correlation, $\mathcal{C}$ to a class of non-Markovian witnesses, $\mathcal{N}$.

The intuition for using contractivity witnesses can be summarized as follows. Markovian processes are contractive, that is, if a map $\mathbb{B}$ is Markovian, then the distinguishability of two input states cannot increase in time, such that 
\begin{gather}\label{contract}
D \left\{ \rho_{1}^S,\, \rho_{2}^S \right\} \geq D \left\{\; \mathbb{B} \left( \rho_{1}^S \right),\, \mathbb{B} \left( \rho_{2}^S \right)\; \right\}. 
\end{gather}
We note that completely positive maps also follow this property. There are many ways to define distinguishability, but they all have similar features. We will focus on the trace distance as a measure of distinguishability, $D\left\{\,\rho_{1},\rho_{2}\,\right\}=\| \rho_1 -\rho_2 \| /2$, where $\| x \| = \mbox{Tr}\sqrt{x^\dagger x}$. The non-Markovianity witness \cite{Breuer:2009cua, BreuerPRL09, Kossakowski:2012wu} is defined in the following way: If there exists times $t,\tau$, with $-\infty \leq t \leq \tau \leq \infty$, such that $D\left\{\rho_{1}^S(t), \rho_{2}^S(t) \right\} \ngtr D \left\{ \;\mathbb{B}_{t,\tau} \left(\rho_{1}^S(t)\right), \mathbb{B}_{t,\tau} \left( \rho_{2}^S(t)\right)\;\right\}$ for some choice of $\rho_{1}^S(t),\rho_{2}^S(t)$, then the process is non-Markovian. In other words, if there is a time segment $[t,\tau]$ where contractivity for some states is violated, then the process is non-Markovian. Note that the test of contractivity depends on the choice of system states, and also on the choice of time parameters $t,\tau$. 

We will use a differential form of contractivity:
 \begin{align}\label{nmwit}
\mathcal{N} =& \frac{d}{dt} D \left\{\, \rho_{1}^S(t),\, \rho_{2}^S(t)\, \right\} \\ =& \lim_{\tau\to t} \frac{\left\| \rho^S_1(\tau) -\rho^S_2(\tau) \right\| - \left\| \rho^S_1(t)-\rho^S_2(t) \right\|} {2(\tau-t)},\nonumber
\end{align}
where $\mathcal{N}$ is time dependent, and a function of the states $\rho_{1}^S,\rho_{2}^S$ at time $t$. This differential form simplifies our analysis as depends on a instant of time $t$. If this $\mathcal{N}> 0$, then the process is not contractive.  In this form, the witness is: If there is a time $t$  at which there are two states $\rho_{1}^S(t),\rho_{2}^S(t)$ that make $\mathcal{N}> 0$, then the process is non-Markovian. This formulation is in agreement with previously published witnesses \cite{Breuer:2009cua, BreuerPRL09} as it captures the idea that, if a process is not contractive at some instance of time, then it is non-Markovian. 

A process can be non-Markovian in general, but seem to be Markovian for some particular time-interval. For this reason, witnesses are often defined that test for all time intervals. We call these witnesses that check for all times $-\infty\leq t\leq \infty$ \emph{strong witnesses}. However, checking for all times has practical difficulties. Other witnesses of non-Markovianity admit this difficulty by testing for non-Markovianity only for a specific time interval \cite{WolfPRL08, Hou:2011dd}. We will now consider another related witness: given some time $t$, if $\mathcal{N}> 0$, then the process is non-Markovian. We will call this a \emph{weak witness}, as it is only checked at a single time $t$. Note that if the weak witness is tested for all possible times, it becomes the strong witness. Since completely positive maps are contractive, and are defined for a specific initial time $t$, this witness also detects violations of complete positivity at time $t$. This weak witness of non-Markovianity is the one we will relate to a witness of initial-correlations. 

Now, we review the important properties of open quantum systems with initial-correlations. Consider states in Eq.~(\ref{map}) with initial $SE$ correlations. States with initial $SE$ correlations are of the form $\rho^{SE}(t)= \rho^S(t)\otimes \rho^E+\chi^{SE}$, where $\chi^{SE}\neq 0$. The initial $SE$ correlations are represented by the correlation matrix $\chi^{SE}$. We remark that this matrix is defined for the initial time $t$. Initial $SE$ correlations, under the right dynamics, can lead to not completely-positive dynamics \cite{pechukas94a, *Jordan:2006p112, *jordan:052110, Rodriguez:2007p123, buzek, Carteret08a, lidarZD, Rodriguez08, Rodriguez11b, embeddingmap, Pernice:2011du, Rodriguez:2008vn, Tan:2011dr}. The complete positivity of the map has been shown to depend on the types of the $SE$ correlations \cite{Rodriguez:2007p123, embeddingmap}, as well as on the specific details of the $SE$ coupling \cite{ShajiDiss, Modi2012a}. Completely positive maps are contractive, following Eq.~(\ref{contract}). It was shown that, for some models, maps coming from $SE$ states with initial-correlation can violate contractivity \cite{ShajiDiss, Modi2012a,PhysRevA.84.042113}. Proposals to detect initial $SE$ correlations have been put forward \cite{Rodriguez:2007p123, LazyStates}. 

A witnesses for initial $SE$ correlations was formulated based on searching for violations of contractivity~\cite{LaineEPL10}. Witnesses of this form rely on the preparation of two system states at the initial time $t$, $\left\{ \rho_{1}^S(t),\rho_{2}^S(t)\right\}$. Although correlations between the system and the environment evolve in time, initial $SE$ correlations are defined as the correlations existing at the  preparation time $t$, and thus the witness $\mathcal{C}$ must be defined for that time. For further work on the more general dynamics of correlations beyond the initial time, please see \cite{Mazzola2012a}. We will discuss later how this choice of a preparation time is similar to the weak non-Markovianity witness described above.

There are many parallels between non-Markovian maps and maps with initial-correlation. A systematic way to treat initial $SE$ correlations and non-Markovian dynamical map has been proposed \cite{Rodriguez08, Rodriguez11b}. In this \emph{canonical map} formalism the dynamics of the initial $SE$ correlations are related to the memory of the non-Markovian process. Also there are witnesses of non-Markovianity and witnesses of initial-correlation that rely on checking for violations of contractivity. These observations suggest a connection between them. We will now find this connection.

\section{State Preparation}

Although the witness proposals for non-Markovianity and initial-correlation both depend on the preparation of two input states, and on use this to test for violations of contractivity, the relationship between these witnesses has not been worked out.  In order to establish this connection, we first have to discuss the role of state preparation in both witnesses.

The concept of state preparation is a systematic way to analyze the limitations imposed on preparing a quantum mechanical state for a quantum process \cite{kuah:042113, *modiinitial}. It is a theoretical representation of the experimental limitations of preparing a quantum states in order to perform an experiment. For some experiments it is necessary to prepare different quantum states. For example, consider two copies of the evolving total state $\rho^{SE}$, each copy evolves identically until time $t$, where each reduced system $\rho^S$ is subject to a different preparation procedure $\mathbb{P}_1^S$ or $\mathbb{P}_2^S$. Such a preparation procedure is a physically realizable map on the system space $S$ \cite{kuah:042113, *modiinitial, Modi2012a}. The action of such a map on the total state is: $\mathbb{P}^S_j \otimes \mathbb{I}^E \left(\,\rho^{SE}(t)\,\right) =\rho^{SE}_j(t)$, where $\mathbb{I}^E$ is the identity map on the environment and $j=\{1,2\}$ is the label to differentiate between the different preparations on each of the copies of the state. Some examples of preparations are unitary transformations on the system, or von Neumann measurements on the system, see references \cite{kuah:042113, *modiinitial, PhysRevA.81.052119, preppap}.

State preparation is particularly important when characterizing a quantum map by means of quantum process tomography \cite{kuah:042113, *modiinitial, PhysRevA.81.052119}. The preparation time $t$ is used to \emph{define} the initial time of the process considered in Eq.~(\ref{map}), and also to define the initial $SE$ correlations. . Although the preparation is performed on the system Hilbert space, it might affect the $SE$ correlations,
\begin{gather}\label{preparation}
\mathbb{P}_j^S \otimes \mathbb{I}^E \left( \rho^{SE}(t) \right) =\rho^{SE}_j(t)=\rho^S_j(t) \otimes \rho^E+ \mu_j^{SE}.
\end{gather}
The first term shows how the reduced system density matrix depends on the preparation procedure. The matrix $\mu_j$ captures how the rest of the $SE$ state at the initial time $t$ depends on preparation procedure. If there are no $SE$ correlations before the preparation, $\chi^{SE}=0$, then it follows that after preparation $\mathbb{P}_j^S \otimes \mathbb{I}^E \left( \chi^{SE} \right) = \mu_j^{SE}=0$. This preparation procedure shows an equivalent way to treat the state preparation to test violations of contractivity, as in Eq.~(\ref{nmwit}), and it can be used to study the witnesses for non-Markovianity  and the witness for initial-correlation on the same footing. Since preparations must be performed at a specific time $t$, in order to test strong witness of non-Markovianity, in principle an experiments would have to be performed with preparations for all times $t$.

\section{Unification of witnesses}

We will use Eq.~(\ref{preparation}) to derive an expression that contains a relationship between the weak witness of non-Markovianity $N$ and the witness of initial correlation $C$. In order to faithfully relate both witnesses, we will first choose the weak non-Markovianity witness that is define at just one time $t$. Later, we will discussed the implications checking for all times $-\infty \leq t \leq \infty$. This time $t$ is the preparation time, and will also correspond to the initial time for the process. 

We start our derivation by using Eq.~(\ref{map}) to expand the non-Markovianity witness $\mathcal{N}$ at a specific time $t$, as in Eq.~(\ref{nmwit}),
\begin{gather}\label{withembedding}
\mathcal{N} = \lim_{\tau\to t} \frac{\left \|  \mbox{Tr}_E \left[ U \left(\, \rho_1^{SE}(t)-\rho^{SE}_2(t )\,\right) U^\dagger\right] \right\| - \left\| \rho_1^S(t) - \rho_2^S(t) \right\| } {2 \left( \tau-t \right)}.\nonumber
\end{gather}
We then rewrite $\rho_j^{SE}(t)$ using Eq.~(\ref{preparation}) and the subadditivity of the trace distance,
\begin{widetext}
\begin{gather}
\begin{array}{ccccc}\label{bothterms}
\mathcal{N} \leq & 
\underbrace{ \lim_{\tau\to t}\;\frac{1}{2} \frac{\left\|  \mbox{Tr}_E \left[ U \left\{ \rho_1^S(t) -\rho^S_2(t) \right\} \otimes \rho^E U^\dagger \right] \right\|-\left\| \rho^S_1(t)-\rho^S_2(t)\right\|}{\tau-t}} & 
+ & \underbrace{ \lim_{\tau\to t}\;\frac{1}{2}\frac{\left\|  \mbox{Tr}_E\left[U\left(\mu_1^{SE}-\mu_2^{SE}\right) U^\dagger\right] \right\|}{\tau-t}}& \\
&\mathcal{M} &  & \mathcal{C} & 
\end{array}.
\end{gather}
\end{widetext}
This is our main result. Since it is such a lengthy expression, we will refer to the terms in the right side of the inequality of Eq.~(\ref{bothterms}) as $\mathcal{M}$ and $\mathcal{C}$, such that: $\mathcal{N}\leq \mathcal{M}+\mathcal{C}$. We will now examine each of these terms independently.

$\mathcal{M}$ is always contractive, $\mathcal{M} \leq 0$. To prove this, we expand, 
\begin{align}
\mathcal{M} =& \lim_{ \tau \to t} \frac{\left\|  \mbox{Tr}_E \left[U\rho^S_1(t)\otimes\rho^E U^\dagger \right]-\mbox{Tr}_E \left[ U\rho^S_2(t) \otimes \rho^E U^\dagger \right] \right\|} {2 \left( \tau-t \right) }\nonumber\\
&-\lim_{ \tau \to t} \frac{\left\| \rho^S_1(t)-\rho^S_2(t) \right\|} {2 \left( \tau-t \right)}.\nonumber
\end{align}
Note that $\mathcal{M}$ can be written as if it came from a completely positive map, $\mbox{Tr}_E\left[U_{t,\tau} \rho^{S}_j(t)\otimes \rho^{E}U^\dagger_{t,\tau} \right]=\mathbb{B}_{t,\tau}\left(\rho^{S}_j(t) \right)=\rho^S_j(\tau)$. By substituting this into $\mathcal{M}$, we can see that it is contractive, as defined in Eq.~(\ref{contract}). This completes the proof that $\mathcal{M} \leq 0$. The term $\mathcal{M}$ can be thought of as the Markovian part of the total evolution. It captures the idea that even a non-Markovian process behave like a Markovian process for some states.

Now we will focus on the term $\mathcal{C}$ in Eq.~(\ref{bothterms}), to explain how it is a witness of initial-correlation. By taking the limit, we obtain,
\begin{align}\label{markov}
\mathcal{C}=\frac{1}{2}\left\|  \mbox{Tr}_E \left[ H^{SE}(t),\; \left(\mu_1^{SE} -\mu_2^{SE} \right) \right] \right\|.
\end{align}
This term depends on initial-correlation for each of the preparations $\mu_1^{SE}$ and $\mu_2^{SE}$, and on the Hamiltonian coupling between the system and the environment $H(t)$. If there are no initial $SE$ correlations, then $\mu_1^{SE}=\mu_2^{SE}=0$. By modus tollens, this term constitutes a witness of initial $SE$ correlations of the form: If $\mathcal{C}> 0$, then there are initial $SE$ correlations (at time $t$). This witness of initial correlations is related to the previously published witness of initial correlations in Eq.~(3) of Ref.~\cite{LaineEPL10}, by means of $\left\| \rho_1^{SE}-\rho_2^{SE}\right\|-\left\| \rho_1^{S}-\rho_2^{S}\right\|\leq \left\| \mu_1^{SE}-\mu_2^{SE}\right\| $  \footnote{More explicitly, we use Eq.~(\ref{preparation}) to write $\mu_1^{SE}-\mu_2^{SE}+\left( \rho_1^{S}-\rho_2^{S}\right)\otimes\rho^E=\rho_1^{SE}-\rho_2^{SE}$, then take the norm, use the triangle inequality, and reorganize the terms to show that $D\left\{ \mu_1^{SE}-\mu_2^{SE}\right\}\geq D\left\{ \rho_1^{SE}-\rho_2^{SE}\right\}-D\left\{ \rho_1^{S}-\rho_2^{S}\right\}$}. Our witness differs from the previously defined witness \cite{LaineEPL10} in that we have been able to extract a Markovian term from it, leaving explicitly a term that depends of on the changes due to the preparation procedure.

\section{Discussion}

We have shown how the  witnesses of non-Markovianity $\mathcal{N}$ is related to a Markovian part $\mathcal{M}$, and to a witness of initial $SE$ correlations $\mathcal{C}$. If $\mathcal{N}>0$ the process is non-Markovian, if $\mathcal{C}> 0$ the process had initial $SE$ correlations, and where  $\mathcal{M}\leq 0$ always. We would like to discuss the implications of the relationships between these witnesses. We now use $\mathcal{M}\leq 0$ to rewrite Eq.~(\ref{bothterms}) as $\mathcal{N}\leq \mathcal{C}$. This implies that, if non-Markovianity is detected by the witness, there also exists initial-correlation. In other words, the weak witness for non-Markovianity $\mathcal{N}>0$ at time $t$ is a sufficient condition for initial $SE$ correlations at time $t$. 

Recall that $\mathcal{N}$, $\mathcal{M}$ and $\mathcal{C}$ were all defined at some time $t$. To make this more explicitly, we rewrite Eq.~(\ref{bothterms}) as $\mathcal{N}(t)\leq\mathcal{M}(t)+\mathcal{C}(t)$. In this form, we can now discuss the implications this relationship has for the strong non-Markovianity witness. The strong version of the witness required to check for  $\mathcal{N}(t)> 0$ for all $-\infty \leq t\leq \infty$. Since completely positive maps are contractive, applying this witness for an interval of time is equivalent to witnessing the property known as P-divisibility \cite{Hou:2011dd,Vacchini:2011cn}. This strong witness requires us to examine both $\mathcal{M}(t)$ and $\mathcal{C}(t)$ for all time $-\infty \leq t\leq \infty$. This is equivalent to having access to infinite copies of $\rho^{SE}$ for each time $t$, and being able to perform preparations $\mathbb{P}_j$ at each time $t$. From this, we can follow the procedure to derive that $\mathcal{M}\leq0$ at each time $t$, and we reach the conclusion that $\mathcal{M}(t)\leq0$ for all $-\infty \leq t\leq \infty$. Thus, $\mathcal{M}(t)$ captures the Markovian part of the evolution at all times. The role of $\mathcal{C}(t)$ at all times needs to be examined carefully, as, by definition, initial-correlation correspond to the correlations at the preparation time $t$. Thus, if $\mathcal{C}(t)>0$ for some time $t$, all we can conclude is that, if a process had been started by preparing states at time $t$, then the process would have had initial-correlation. This highlights the fact that initial-correlation is a property of one time, while full non-Markovianiaty is a property of all times.

\section{Conclusion}

We have established a connection between a witness of initial-correlation to a witness of of non-Markovianity and not-completely positive maps. This connections exploits the property of contractivity of Markovian processes and of completely positive maps at an instance of time. This work suggest how to establish connections between research of dynamical maps with initial system-environment correlations and properties of non-Markovian open quantum systems at an instance of time. We conjecture that this interplay will hold for other types of witnesses, and will suggest new ways to detect systems with initial system-environment correlations.

\section{Acknowledgements}

LM gratefully acknowledge Andrea Smirne and Bassano Vacchini for invaluable discussions. LM and CARR thank the Centre for Quantum Technology, for the kind hospitality during the early stages of this work. We acknowledge the Magnus Ehrnrooth Foundation and the UK EPSRC (EP/G004759/1) for financial support. KM is supported by the John Templeton Foundation, the National Research Foundation, and the Ministry of Education of Singapore.

\bibliography{nonmarkov}

\end{document}